\documentclass[aps,prl,preprint]{revtex4-2}
\usepackage{graphicx}

\begin{document}

\title{Is a  Dissipative System Always a Gradient  System or a Gradient Like System?}


\author{Rafael Rangel$^{1}$}
\email[]{rerangel@usb.ve}
\
\affiliation{$^{1} $ Departamento de F\'{i}sica,  Universidad Sim\'{o}n Bol\'{i}var, A.P. $89000$, Caracas 1080-A, Venezuela}


\date{\today}
\begin{abstract}
 We find that  to  a given  dissipative system  a p=1 differential form  can be associated which has  a  general decomposition into 
 a potential term and a non-potential residual part. If the residual part is absent the form is closed and the system is gradient system or gradient like. If it is non-closed, in the differential form approach,  it remains non-closed under a variable  change of  coordinates, i.e., the system  is not  a gradient one or a  gradient like in any coordinate system. On the other hand, there are claims that a potential  should always exists, i.e., the class of dissipative systems and the the class of gradient systems should  coincide.   We fix this conundrum  by introducing   a generalized  change of coordinates that aims a  transformation to a gradient system or a gradient like system. The condition of being closed in the new coordinates of  a certain, through the generalized change of coordinates  defined  differential form,  results in  a nonlinear differential equation  together with a consistency condition. We give examples of physical systems where an analytical solution for the transformation can be found, and hitherto, the potential, but even   when the potential is not accessible analytically, we find  that it always exist, and therefore we give in principle an affirmative answer to the defining question of this work.
 Our findings  removes loopholes in the question if a potential  may exist but it is not known.
\end{abstract}

\pacs{74.81.Fa,05.10.Gg}

\maketitle


\section{ Introduction.}

Dynamical systems represented  by a vector $\vec{x}$ in $\in R^{N}$, and whose evolution is   given by  stochastic equation(SDE):
\begin{equation}
\dot{\vec{x}}=\frac{d\vec{x}}{dt}=\vec{f}(\vec{x}) + \vec{g}(\vec{x})\zeta(\vec{x},t)
\label{SODE}
\end{equation}
 constitutes a universal matter of study \cite{BerntOksendal}. Here each of the functions $f(\vec{x})$ and  $g(\vec{x})$ has sufficient well behavior such that ODE to have solutions, and the function $\zeta(t)$ represent an stochastic process (noise)\cite{CramerLeadbetter}taken to be a Markov process. 
This equation without the  noise term, has been intensively studied for the nature of the dynamics in particular, when the system is dissipative\cite{GuckenheimerHolmes}, \cite{StephenWiggins}\cite{HCartanI}. Another issue is
concerned with the existence of Lyapunov functions and its role in the classification of the dynamics\cite{MSanMiguelRToral}. 
If there exist  a scalar function $V(\vec{x})=V(x_{i})$, $ i=1,..N$, which is  usually called the potential,  such that it is bounded from below, and stationary on the limits set of the dynamical systems \cite{GuckenheimerHolmes} and such that:
\begin{equation}
\dot{\vec{x_{i}}}=f_{i}(\vec{x})=-\frac{\partial V(\vec{x})}{\partial x_{i}}\Rightarrow \frac{dV(\vec{x})}{dt}=-\sum_{i=1}^{N}\frac{\partial V(\vec{x})}{\partial x_{i}}\frac{\vec{x}_{i}}{dt}=\sum_{i=1}^{N}\left(\frac{\partial V(\vec{x})}{\partial {x_{i}}}\right)^{2} \le 0,
\label{Lyapunov1}
\end{equation}
the potential  is called a Lyapunov function for the system \cite{GuckenheimerHolmes}. One can consider a more general situation given in which in the governing dynamics a residual vector function 
${\vec{\it{v}}}$ appears, in addition to a positive definite matrix $\mathcal{S}(\vec{x})$:
\begin{eqnarray}
\dot{\vec{x_{i}}}=\vec{f}_{i}(\vec{x}):=-\sum_{j=1}^{N}\mathcal{S}_{ij}\frac{\partial \it{V}(\vec{x})}{\partial x_{j}} + \vec{\it{v}}_{i}(\vec{x}),\nonumber\\
{\vec{\it{v}}}\cdot\nabla\vec{V}=0
\label{Lyapunov2}
\end{eqnarray}
i.e.,  the vector functions  ${\vec{\it{v}}}$  and   $\vec{\nabla} V(\vec{x})$ are orthogonal in the configuration space $U$  $\subset R^{N}$ (orthogonality condition). In this case  the potential $\it{V}(\vec{x})$  still behaves as a Laypunov function, i.e., $\frac{dV(\vec{x})}{dt}\le0$.
The orthogonality condition can be written, by directing rewriting  it in the form:
\begin{eqnarray}
\left(\vec{f}(\vec{x})+ \mathcal{S}\vec{\nabla} V(\vec{x})\right)^{T}\cdot\vec{\nabla} V(\vec{x})=\vec{f}^{T}(\vec{x})\vec{\nabla} V(\vec{x})+(\vec{\nabla} V(\vec{x}))^{T}\mathcal{S}\vec{\nabla} V(\vec{x})=0
\label{OrthogonalityCondition}
\end{eqnarray}
Previous equation, giving the  orthogonality condition and the function $\vec{f}(\vec{x})$,  can be though as an equation for determining  $V(\vec{x})$ \cite{DezaRRWioHS}. 
This is Graham and Tel\cite{GrahamTel} approach, where the existence of  a globally defined potential function with respect to the positive define matrix $\mathcal{S}$ is defined.
The residual part ${\vec{\it{v}}}$, does not participate in the relaxation dynamics.
When considering Equ.(\ref{SODE}) for the special case of a Gaussian random processes, i.e.:
\begin{eqnarray}
\dot{\vec{x}_{i}}=\vec{f}_{i}(\vec{x}) + \sum_{j}^{N}\vec{g}_{ij}(\vec{x})\zeta_{j}(t);  \hspace{1cm}<\zeta_{i}(t)\zeta_{j}(t)>=2\epsilon\delta(t-t^{\prime}),\nonumber\\
\vec{f}_{i}=-\sum_{j=1}^{N}\mathcal{S}_{ij}\frac{\partial \it{V}(\vec{x})}{\partial x_{j}} + \vec{\it{v}}_{i}\hspace{6.5cm}\nonumber\\
\label{SODE1}
\end{eqnarray}
The Ansatz for $\vec{f}_{i}$ in previous equation fulfills the orthogonality condition.  Now, the question for finding the potential (see \cite{WioHSDezaRR} for a review), when considering the associated Fokker-Planck  to Equ.(\ref{SODE1}), and other 
assumptions, namely,  the validity of the fluctuation dissipation relation for the case when $\mathcal{S}$ is a constant matrix,  when  $\mathcal{S}=gg^{T}$, and the validly of the orthogonal relation and the divergenz free of $\vec{v}$, then there exist a stationary distribution $P_{stat}$ for the associated Fokker-Plank-equation given by in the small noise limit (lim $\epsilon \rightarrow 0$):
\begin{equation}
P_{stat}=Nexp\left(-\frac{V(\vec{x})}{\epsilon}\right)
\label{StationaryDistribution}
\end{equation}
from which one could obtain the potential if $P_{stat}$  could be independently  obtained, as the limit: 
\begin{equation}
V(\vec{x})=lim_{\epsilon \rightarrow 0} (-\epsilon \ln{P_{stat}})
\label{GrahamPotential}
\end{equation}
 also known as Graham potential. Graham and Tel \cite{GrahamTel} have pioneered the question for the general conditions for  a potential to exists for dissipative dynamical systems.  They have shown that both ways of reasoning  are equivalent, i.e., the potential obtained from Equ.(\ref{OrthogonalityCondition}) and the one from Equ.(\ref{GrahamPotential}). 
They also study the concrete question of sufficient conditions for  the existence of a potential, given a function $f(\vec{x})$.
The method bases on solving the question of the integrability of a formal associated of a Hamilton systems to the Equ.(\ref{ODE}). However, we demonstrate in the appendix, that their scheme is just the condition for certain diff-form to be closed \cite{WioHSDezaRR}, or equivalently, to posses an integrating factor. Also,  we show below that for every dissipative systems there is a decomposition quantitatively similar to equation Eqn.(\ref{Lyapunov2}).  
Remarkably,  the question if a given dissipative systems is a gradient like systems also find relevance in mathematics.
 There are claims \cite{RalphChillEvaFasargova} that  with high probability one  will not be able to avert that question. However, the authors Chill and Fasarngova  do not shrug off the importance of the question, on the contrary, they express hope by saying that they seem to see gradient systems every where and reaffirm the importance of the question by saying that they hail a gradient system whenever they see one. But their main inquires  are how to show that a dissipative systems is a gradient systems and how
 to find the potential when one feels  it exist?.\\
In this work we analyse this question\cite{RalphChill}, specifically  for systems like:
\begin{equation}
\beta_{c_{j}}\ddot {x}_{j}+\dot {x}_{j}+g_{j}(x_{1},\ldots,x_{N+1}) + \zeta_{j}(t)=0; \hspace{0.5cm} j=1,\ldots,N+1
\label{ODE}
\end{equation}
These are in fact a sufficient general systems which contain interesting limits. In this system, the $x_{j}$ variables defines a vector in $\vec{x}\in R^{N+1}$, and  the functions $\vec{g}=g_{j}(x_{1},\ldots,x_{N+1})$ defines the vector function $\vec{g}\in R^{N+1}$, these functions are assumed to be at least twice continuous differentiable, such that existence and uniqueness of classical solutions the ODE system is guarantee \cite{HirschSmaleDevaney}.
In addition, the last term represents a white noise term, ever present in physical systems, therefore we are dealing with stochastic differential equations.
 The parameters $\beta_{cj}$ is a measure of an inertia like concept in physics.
In cases where the noise term can be neglected, justified for example at low enough temperatures,  one can analyze the system by the flow properties of its  associated first order system:
\begin{eqnarray}
 \vec{\beta}_{c}\dot{\vec{x}}= \vec{\bar{x}}\nonumber\\
\dot {\vec{\bar{x}}}= -\frac{1}{\beta_{c}}\vec{\bar{x}}-\vec{g}(\vec{x})
\label{firstodersystem}
\end{eqnarray}
 where $\vec{\beta}_{c}=\beta_{1},...., \beta_{N+1}$,  $\vec{x}=(x_{1},..,x_{N+1})$, \hspace{0.08cm}$\vec{\bar{x}}=(x_{N+1},..,x_{2(N+1)})$ are vectors $\in R^{N+1}$ \cite{GuckenheimerHolmes}\cite{StephenWiggins}. If there exist a function , such that $\vec{\nabla} V(\vec{x})=\vec{g}(\vec{x})$, we say the system is gradient like, but note that It is not a gradient system.
 Here we are in the case of a  the dynamical system which  is phase space contracting like the systems treated in \cite{GrahamTel} and write for it for short: $\dot{x_{i}}=X_{i}$.
 For negligible inertia, one has the over-damped limit: $ \dot {\vec{x}}= -\vec{g}(\vec{x})$. 
 On the other hand, if the dissipative term (second term in Equ.(\ref{ODE})), or second  term in the second equation in Equ.(\ref{firstodersystem})) is negligible as compared with the inertial term, and there is a potential function $V(x_{1},..,x_{N+1})$, 
the systems is Hamiltonian with Hamilton function given by $\mathcal{H}_{P}(\vec{x},\vec{y})=\left(\beta_{c}\sum_{i=1}^{N+1}\frac{1}{2}y_{i}^{2} + V(x_{1},..,x_{N+1})\right)$:\nonumber\\
\hspace{0.2cm}  $\beta_{c}\dot{\vec{x}}= \nabla_{\vec{x}}\mathcal{H}_{P}(\vec{x},\vec{y})$; \hspace{0.2cm} $\dot{\vec{y}}=- \nabla_{\vec{x}}\mathcal{H}_{P}(\vec{x},\vec{y})$.
\section{Differential Forms and the existence of a Potential.}
 Differential forms   \cite{HansSamelson}\cite{HCartanII}\cite{HFlanders} have being  used in  expressing in a natural way fundamental theories like the Maxwell equations for electromagnetism  or the Einstein general theory of relativity  \cite{WThirringI}\cite{WThirringII} (see  \cite{MaxwelleqtsExample}). Differential forms   are basic for defining the symplectic  structure of classical Hamiltonian  mechanics\cite{VIArnold}.
For our purposes we introduce  the essentials needed in the sequel, and adhere to the  conviction in
 \cite{DavidBetounes} that  this rather short introduction will position the reader in catching the main points  of our approach.\\
 Let $ x \in \mathcal{U} \subseteq R^{N+1}$, and $\omega(x)$  a function to the space of the p-linear alternating functions $\mathcal{A}_{p}(R^{N+1},R)$ , i.e.,  functions from  $R^{p}\rightarrow R$, and the meaning of  alternating just below.
Naming the vector space of all differential forms by $\Omega^{n}_{p}(R^{N+1},R)$ which are of class $C^{n}$, $\omega(x)\in \Omega^{n}_{p}$ means  $\omega(x,\xi_{1},....\xi{p})\in R$,  and  alternating in the variables $\xi_{j}$ means  $\epsilon(\sigma)\omega(x,\xi_{\sigma(1)},....\xi_{\sigma(p)})=\omega(x,\xi_{1},....,\xi_{p})$, where $\epsilon(\sigma)$ denotes the signature of the permutation of the variables $\xi_{j}$\cite{HCartanII}.
Now naturally the question of differentiation appears, and the derived function\cite{HCartanI} of $\omega(x,\xi_{1},....,\xi_{p})$ denoted by $\omega\prime$  is a functional from $U$ into the space  
$\mathcal{L}(R^{N+1};\mathcal{A}_{p}(R^{N+1},R)$, i.e.,  the space of functionals from $R^{N+1}$ into $\mathcal{A}_{p}(R^{N+1},R)$, i.e., \space $\omega\prime(x;\xi_{0})(\xi_{1},..,\xi_{p})\in R $, $(\xi_{0},\xi_{1},..,\xi_{p})\in R^{N+1}$. 
However, $\omega\prime(x;\xi_{0})(\xi_{1},..,\xi_{p})$ is not an alternating function of $(\xi_{0},\xi_{1},..,\xi_{p})$, it is an alternating function of the $(\xi_{j}, j=1,..p$, and linear in the $(\xi_{0})$.
To consistently stay within the  multi-linear  alternating   class of functions, one apply the same procedure employed for the consistently definition of  multiplication of two alternating functions, i.e.,  one ($f(x_{1}, ..., x_{p})\in R)$ in
 $\mathcal{A}_{p}(R^{N+1},R)$  and  another ($g(x_{p+1}, ..., x_{p+q})\in R)$ in $\mathcal{A}_{p}(R^{N+1
 },R)$.
The  canonical form realized via  a bilinear mapping $\phi_{p,q}: \mathcal{A}_{p,q}(R^{p},R)\rightarrow  \mathcal{A}_{p+q}(R^{p},R)$, such that the product: $\it{h}:=f(x_{1}, ..., x_{p})g(x_{p+1}, ..., x_{p+q})\rightarrow \tilde{\it{h}}=\sum_{\sigma}\epsilon(\sigma)(\sigma\it{h})$, and the summations is over all permutations of the set $\{1,....p+q\}$ such that $\sigma(1)<,....,<\sigma(p)$, and $\sigma(p+1)<,....,<\sigma(p+q)$.
Accordingly,  summarizing we have then: 
\begin{widetext}
\begin{eqnarray}
x \in U, \hspace{0.3cm} \omega(x)\Rightarrow \mathcal{A}_{p}(R^{N+1},R)\nonumber\\
\omega^{\prime}(x): \mathcal{U}\Rightarrow \mathcal{L}(R^{N+1};\mathcal{A}_{p}(R^{N+1},R))\nonumber\\
\omega^{\prime}(x;\xi_{0})(\xi_{1},..,\xi_{p})\in \mathcal{A}_{1,p}(R^{N+1},R) \stackrel{d\omega}{\longrightarrow} \mathcal{A}_{1+p}(R^{N+1},R) \nonumber\\
d\omega(x;\xi_{0},\xi_{1},..,\xi_{p})=\sum_{i=1}^{p}(-1)^{i}\omega^{\prime}(x;\xi_{i})(\xi_{1},..,\tilde{\xi}_{i},.....\xi_{p})\nonumber\\
d(d\omega(x;\xi_{0},\xi_{1},..,\xi_{p}))=0
\label{diffformsdefinitions}
\end{eqnarray}
\end{widetext}
The tilde over $\xi_{i}$ means it is omitted from  the variables list. This explicit expression for  the d-operator $d\omega$ is a result of the definition above for product of differential forms.
Now we must consider another  operation on  differential forms, i.e., the change of variable procedure\cite{DavidBetounes}.
Given a mapping $\phi\in C^{n+1}$ from a subspace $U^{\prime}\in R^{N+1}$ to another $U \in R^{N+1}$, and given that one has differential p-forms defined in $U$, the mapping permits to define
differential forms on $U^{\prime}$. We assume that the inverse mapping, $\phi^{-1}$ exist and  $\in C^{n+1}$:
\begin{widetext}
\begin{eqnarray}
\vec{x} \in U, \hspace{0.3cm} \omega(x)\Rightarrow \mathcal{A}_{p}(R^{N+1},R)\nonumber\\
\phi: U^{\prime} \rightarrow U,\hspace{0.2cm} \vec{y} \in U^{\prime},\hspace{0.1cm} x_{i}=\phi_{i}(y_{1},..,y_{N+1}), \hspace{0.3cm} \phi^{\star}(\omega): U^{\prime} \stackrel{\phi^{\star}}{\longrightarrow}  \mathcal{A}_{p}(R^{N+1},R) \nonumber\\
(\eta_{1},..,\eta_{p}) \in R^{N+1},\hspace{0.3cm}(\phi^{\star}(\omega))(y;\eta_{1},..,\eta_{p}):=\omega(\phi(y);\phi^{\prime}(y)\cdot\eta_{1}),.....,\phi^{\prime}(y)\cdot\eta_{p})\nonumber\\
d(\phi^{\star}(\omega))=\phi^{\star}(d\omega)
\label{changeofvaribalestrans}
\end{eqnarray}
\end{widetext}
The change  of coordinate is a linear mapping in $\Omega^{n}_{p}(R^{N+1},R)$, the last line express the core of the consistency of the theory, the value of the $d\omega$ is independent of the coordinates systems, i.e., invariant under change of variables\cite{HFlanders}.  Another remarkably aspect of the theory is Poincare $\grave{}s$ theorem \cite{Note1}.
Assume from now on that the set $U$ is starlike with respect one of its points  say the point a, if the for all x,  $\in U$, the segment $(1-t)a + tx$, $0\le t \le 1$   is contained in U \cite{WThirringI}, in particular a convex set is starlike with respect to any of its points. One can take in full generality $a=0$,  the zero of the vector space $R^{N+1}$.
We now consider particular mapping that was first  proposed in \cite{AndreWeil} and define the operator $k(\omega)$, that acts on differential  forms: 
\begin{widetext}
\begin{eqnarray}
(\xi_{1},..,\xi_{p-1}) \rightarrow \omega(tx;x,\xi_{1},..,\xi_{p-1}),\hspace{0.2cm} tx\in U\nonumber\\
k(\omega)(x,\xi_{1},..,\xi_{p-1})=\int\limits_{0}^{1}t^{p-1}\omega(tx;x,\xi_{1},..,\xi_{p-1})dt\nonumber\\
k: \Omega^{n}_{p}(R^{N+1},R) \rightarrow \Omega^{n}_{p-1}(R^{N+1},R)\nonumber\\
\label{kopt}
\end{eqnarray}
\end{widetext}
The integral is understood in the usual Riemann sense.
The mapping is particular in which the argument of the differential  form $\omega$ for which  it is defined in $U$ is the same as one of the  first vector where it must be evaluated to obtain finally its value $\in R$. Therefore,  the operator $k(\omega)$ is alternating  in the variables $(\xi_{1},..,\xi_{p-1})$.  In addition, the differentiability assumptions on  $(\omega)$ permits to differentiate under the integral sign,  
then the important result is obtained \cite{HCartanII}, \cite{WThirringI}:
\begin{widetext}
\begin{eqnarray}
d(k(\omega)) + k(d\omega)=\omega
\label{dkopt}
\end{eqnarray}
\end{widetext}
If $d(\omega)=0$,  i.e.,  the differential form $\omega$ is a closed form by definition, then $d(k(\omega)))=\omega$, this is Poincare$\grave{}$s Theorem \cite{Note1}
in which case one names $\omega$ an exact form. For starlike sets U, the closed condition is necessary and sufficient for a form to be exact. The decomposition of $\omega$ is unique and invariant under change of variables( \cite{MaxwelleqtsExample}).
For our purposes, in the sequel, we consider  the canonical $p=1$-differential form $\omega$  \cite{HCartanII}. In this case, the operator $k$ in Equ.(\ref{kopt}) 
transforms a $p=1$ differential form $\omega$ into a function, that function will represent the potential part of $\omega$ or  just the potential when it ia closed form.
then, 
 using the g-functions  from  Equ.(\ref{ODE}) we  give  explicitly  the condition such that it is a closed  using the explicit form for $d(\omega)$ in Equ.(\ref{diffformsdefinitions}) for $p=1$:
\begin{widetext}
\begin{eqnarray}
\omega:=\mathcal{G}=\sum_{j=1}^{N+1} g_{j}(x_{1},\ldots,x_{N+1})dx_{j};\hspace{0.5cm} \mathcal{G}\in \Omega^{n}_{1}(R^{N+1},R),\nonumber\\
\mathcal{G}(\vec{x},\vec{\Xi})=\sum_{j=1}^{N+1} g_{j}(x_{1},\ldots,x_{N+1})\xi_{j}\nonumber\\
d\mathcal{G}(\vec{x},\xi_{1},\xi_{2})=(\omega^{\prime}(\vec{x};\xi_{1}))(\xi_{2})-(\omega^{\prime}(\vec{x};\xi_{2}))(\xi_{1})=0 \hspace{0.5cm} \Rightarrow\nonumber\\
 \left [g_{j}(\vec{x}),\partial{x_{l}}\right]:=\frac{\partial{g_{j}(\vec{x})}}{\partial{x_{l}}}=[g_{j}(\vec{x}),\partial{x_{l}}]^{T}=[g_{l}(\vec{x}),\partial{x_{j}}]
\label{dynamicssystemdiffform}
\end{eqnarray}
\end{widetext}
The last conditions says that  the matrix  $[g_{j}(\vec{x}),\partial{x_{l}}]_{j,l}$ of the partial derivatives is symmetric.
 In the sequel we prove previous statement for our $p=1$ diff-form.
From Equ.(\ref{dynamicssystemdiffform}) and the definition Euq.(\ref{kopt}) we get:
\begin{eqnarray}
k(d\omega)((\vec{x})\cdot \xi_{1}=\int\limits_{0}^{1}t[(\omega^{\prime}(\vec{tx};\vec{x})\cdot\xi_{1})-(\omega^{\prime}(\vec{tx};\xi_{1})\cdot\vec{x})]dt=\nonumber\\
\int\limits_{0}^{1}t\sum_{i,j;i\neq j}[\frac{\partial{g_{i}(\vec{t\vec{x}})}}{\partial{x_{j}}}-\frac{\partial{g_{j}(\vec{t\vec{x}})}}{\partial{x_{i}}}](x_{i}\xi^{j}_{1}-\xi^{j}_{1}x_{i})dt\nonumber\\
d(k(\omega))(\vec{x}) \cdot \xi_{1}=\int\limits_{0}^{1}[t(\omega^{\prime}(\vec{tx};\xi_{1})\cdot\vec{x})+ \omega(\vec{tx})\cdot\xi_{1}]dt\hspace{1.4cm}\nonumber\\
\label{explizitcalculation}
\end{eqnarray}
Then, we have the following facts:
\begin{eqnarray}
\frac{\partial{g_{i}(\vec{t\vec{x}})}}{\partial{x_{j}}}=\frac{\partial{g_{j}(\vec{t\vec{x}})}}{\partial{x_{i}}} \iff d\omega =0 \iff
[(\omega^{\prime}(\vec{tx};\vec{x})\cdot\xi_{1})=(\omega^{\prime}(\vec{tx};\xi_{1})\cdot\vec{x})] \Rightarrow \nonumber\\
t(\omega^{\prime}(\vec{tx};\xi_{1})\cdot\vec{x})+ \omega(\vec{tx};\vec{x})\cdot\xi_{1}=(t\omega^{\prime}(\vec{tx};\vec{x})+ \omega(\vec{tx};\vec{x}))\cdot\xi_{1} =\frac{d(t\omega(t\vec{x}))}{dt}\cdot \xi_{1} \Rightarrow\nonumber\\
d(k(\omega))(\vec{x})=\int\limits_{0}^{1}\frac{d(t\omega(t\vec{x}))}{dt}=\omega(\vec{x})\nonumber\\
\label{dwZero}
\end{eqnarray}
See \cite{Note5}.
Although we are dealing  with  differential forms, the last relation can be considered as an equality of functions in $\mathcal{A}_{1}(R^{N+1},R)$. This procedure is central to our reasoning below where we are pursuing the functions defining the p=1 diff=form. 
As a result, we have that:
\begin{eqnarray}
k(\omega)(\vec{x})=\int\limits_{0}^{1}\omega(\vec{tx};\vec{x})dt=\int\limits_{x_{0}=0}^{\vec{x}}\omega(\vec{x})dx=\sum^{N}_{i=1}\int\limits_{0}^{1}x_{i}g_{i}(t\vec{x})dt\nonumber\\
\Rightarrow \frac{\partial{k(\omega)(\vec{x})}}{\partial{x_{i}}}=g_{i}(\vec{x})\nonumber\\
\label{Dasdingansich}
\end{eqnarray}
the integral is along  a linear segment from $x_{0}$ to $x$ \cite{HCartanII}. We name the second line in previous equation the consistency condition. Now,  $k(\omega)(\vec{x})$ defines a function in $U$, and the integral of $\omega(\vec{x})$  around  any  class $C^{1}$ loop $\gamma $ is zero\cite{HCartanII}, in fact, $d\mathcal{G}= 0 \iff \int\limits_{\gamma}\omega (\vec{x})=0$. Equ.(\ref{Dasdingansich})  defines  the potential.
When   $d\mathcal{G}\neq 0$, i.e., $\mathcal{G}$ is not closed, the first term  in Eqn.(\ref{dkopt}) defines a closed form, ($dd(k(\omega))=0)$, and  and a non-coed form called 
 the residual part $k(d\mathcal{G})$. In other words, the lack of the
nice property of the equality of the cross  derivatives that straightforwardly conducted us to the potential function,  now makes the a residual part according to Eqn.(\ref{explizitcalculation}).  In either case, the matrix of the partial derivatives $ \left [g_{j}(\vec{x}),\partial{x_{l}}\right]$ 
is the main object to analyze. No coordinate transformation bring us outside this relation, whose demonstration is now straightforward as follows:
Consider  the sum of both terms in Eqn.(\ref{explizitcalculation}),
\begin{eqnarray}
k(d\omega)((\vec{x})\cdot \xi_{1}+d(k(\omega))(\vec{x}) \cdot\xi_{1}=\int\limits_{0}^{1}t[(\omega^{\prime}(\vec{tx};\vec{x})\cdot\xi_{1})-(\omega^{\prime}(\vec{tx};\xi_{1})\cdot\vec{x})]dt+\nonumber\\
\int\limits_{0}^{1}[t(\omega^{\prime}(\vec{tx};\xi_{1})\cdot\vec{x})+ \omega(\vec{tx})\cdot\xi_{1}]dt=\int\limits_{0}^{1}[t(\omega^{\prime}(\vec{tx})\cdot\vec{x}) + \omega(\vec{tx})dt]\cdot\xi_{1}=\nonumber\\
\int\limits_{0}^{1}\frac{d(t\omega(t\vec{x}))}{dt}\cdot\xi_{1}=
\omega(\vec{x})\cdot\xi_{1}=\mathcal{G}(\vec{x},\vec{\Xi});\hspace{2cm}\nonumber\\
\label{GeneralizedPoincare}
\end{eqnarray}
i.e., the cancelation of  the first term in the third line and second term in first line in Equ.(\ref{explizitcalculation})  permits the same  manipulations given in the second line in Equ.(\ref{dwZero}) \cite{Note2}\cite{MSanMiguelRToral}.
From  Equ.(\ref{changeofvaribalestrans}),  the non-closed condition can not be cure by any change of variables. Then, from Eqn.(\ref{dwZero}), given a dissipative  systems the first thing to check pursuing the existence of a potential function  is the equality of the cross-derivatives of the associated p=1 diff-form.  When  this is not the case,  the following ideas are pertinent.
\subsection{Interpretation of the dynamics in term of Differential Forms}
Now, a central point in our approach  is  to interpret the differential equation Equ.(\ref{firstodersystem}) in terms of differential forms, fort his purpose we define:
\begin{eqnarray}
\dot{\vec{y}}=Y(\vec{y})\Rightarrow \Psi:=d\vec{y}-Y(\vec{y})dt
\label{Totaldiffeqn}
\end{eqnarray}
The p=1 differential form $\Psi$, defined in a set $U\subseteq R^{n}XR$.  When $Y_{i}(\vec{y})$ $\in C^{1}(\vec{y})$, Frobenious theorem \cite{HCartanII} establish the existence of a unique solution $\vec{y}_{s}(t,\vec{y})$,  the one that annihilates the  differential form $\Psi$, i.e.,  $\dot{\vec{y}}_{s}(t,\vec{y})dt-Y(\vec{y}_{s})dt=0$, 
and so therefore, by considering  equality between  elements in $\Omega^{n}_{1}(R^{N+1},R)$, 
 we have  within the differential forms formalism the validity of following relation\cite{Note3}(see Eqn.(\ref{firstodersystem})):
\begin{eqnarray}
 \beta_{c}d\vec{x}= \vec{y}dt\hspace{11.7cm}\nonumber\\
d\vec{y}= -\frac{1}{\beta_{c}}\vec{y}dt-[d(k(\mathcal{G}))(\vec{y})+  k(d\mathcal{G})(\vec{y})]dt;
\hspace{5.73cm}\nonumber\\
d(k(\mathcal{G}))(\vec{y})+  k(d\mathcal{G})(\vec{y})=
\int\limits_{0}^{1}t(\mathcal{G}^{\prime}(\vec{tx})\cdot\vec{x})dt + \int\limits_{0}^{1}\mathcal{G}(\vec{tx})dt\hspace{4.0cm}\nonumber\\
\label{firstodersystemwithdkopt}
\end{eqnarray}
 The whole bracket  make sense as a function, i.e., the considered equality of elements
in $\Omega^{n}_{1}(R^{N+1},R)$ works, i.e., we always have in mind the canonical representation of diff-form, and for conclusions only the defining function of the diff-form matters.
We conclude, by comparing  Equ.(\ref{firstodersystemwithdkopt}) with Equ.(\ref{Lyapunov2}), this the most general possible case for any systems defined by Equ.(\ref{firstodersystem}), and that the orthogonality condition is a particular one. 
Any  suitable change of variables $\phi_{i}(y_{1},..,y_{N+1})$ that  express the same SOD 
in other coordinates \cite{OdinetteReneeAbib} leaves invariant the structure of  Equ.(\ref{firstodersystemwithdkopt}).
As an example, we take  the Lorenz equations\cite{GuckenheimerHolmes}. With  $\vec{\mathbf{y}}=(x,y,z)$$\in R^{3}$, the vector function $\vec{\mathbf{G}}=(g_{1}(\vec{\mathbf{y}}),g_{2}(\vec{\mathbf{y}}),g_{3}(\vec{\mathbf{y}}))$, with 
 ($g_{1}(\vec{\mathbf{y}})=\sigma(y-x),\hspace{0.2cm} g_{2}(\vec{\mathbf{y}})=\rho x-y-xz, \hspace{0.2cm} g_{3}(\vec{\mathbf{y}})=-\beta z+xy)$,
  and the parameters $\sigma,\rho,\beta$ are $>0$.
 For the dissipative system $\dot{\vec{\mathbf{y}}}=\vec{\mathbf{G}}$,  one easily verifies  that the matrix function of the  partial derivatives is  not symmetric,
 in fact one finds $\partial g_{i}(\vec{\mathbf{y}})/\partial \mathbf{y}_{j} \neq \partial g_{j}(\vec{\mathbf{y}})/\partial \mathbf{y}_{i}$, $1\leq i,j \le  3$.
 Therefore,  for  the associated $p=1$ differential form  $\mathcal{G}(\vec{\mathbf{\mathbf{y}}})$(Equ.(\ref{dynamicssystemdiffform})), 
  $d\mathcal{G}(\vec{\mathbf{\mathbf{y}}})\not=0$, and therefore, there  is not a potential  in the sense of Equ.(\ref{Lyapunov1},\ref{Dasdingansich}). In addition, we have
  $\vec{\mathbf{G}}\cdot\nabla\times\vec{\mathbf{G}}\neq 0$ and  there is no sense in trying the Graham scheme (see appendix, Eqn.(\ref{factorexistencecondition})), \cite{WioHSDezaRR}\cite{DezaRRWioHS}, notwithstanding,   the universal  relation applies(see\cite{DominicEdelen}, theorem 5-3.1, where the operator k is denoted by $\mathcal{H}$, the Homotopy operator, see also theorem 5-6.1, and the corollaries  5-6.1-3):
\begin{eqnarray}
\omega=\omega_{exact}+ \omega_{antiexact}:=d(k(\mathcal{G}))(\vec{\mathbf{y}})+ k(d\mathcal{G})(\vec{\mathbf{y}})\Rightarrow\nonumber\\
d(\omega)=d(d(k(\mathcal{G}))(\vec{\mathbf{y}})) + d( k(d\mathcal{G})(\vec{\mathbf{y}}))=d (k(d\mathcal{G})(\vec{\mathbf{y}}))=d\mathcal{G}(\vec{\mathbf{y}})\nonumber\\
k(\omega)=k(d(k(\mathcal{G}))(\vec{\mathbf{y}}))+ k(k(d\mathcal{G})(\vec{\mathbf{y}}))=k(d(k(\mathcal{G}))(\vec{\mathbf{y}}))=k(\mathcal{G})\nonumber\\
\label{omegaParts}
\end{eqnarray}
The descomposition of $\omega$ is such, that the first term is in the kernel  of the operator d, and the second term in the kernel of the operator k.
From visual comparison of previous equation with Equ.(\ref{Lyapunov2}), we define in full generality, the first term  $\omega_{exact}:=dk(\mathcal{G}))(\vec{y})$ as the potential part, it is a closed form ($dd(k(\mathcal{G}))(\vec{\mathbf{y}})=0)$ and therefore ($d(k(\mathcal{G}))(\vec{\mathbf{y}})$ is an exact form, therefore there exist a function 
$\mathcal{G}_{pot}:=k(\mathcal{G}))(\vec{\mathbf{y}})$,  such that $d\mathcal{G}_{pot}=\omega_{exact}$. In analogy, 
  we name the second term, $\omega_{antiexact}:=k(d\mathcal{G})(\vec{y})$ 
as the residual part, it is invariant under the action  of  the operator $(dk)$ and belongs to the kernel of the operator k \cite{Note4},  $k(k(d\mathcal{G}))=0$ , and therefore
there is a  p=1+1 diff-form, namely and explicit  $d\mathcal{G}$.
In fact, \cite{DominicEdelen}, $\omega_{exact}$ and $\omega_{antiexact}$ belong to sets whose intersection is the null element in the set of the  $p=1$ differential forms, this property can  be considered as an orthogonality like relation, and this means, the representation through exact and antiexact forms is unique.
For the Lorentz system, es is an easy task to calculate $\mathcal {G}_{exact}$  and $\mathcal {G}_{antiexact}$. 
But the message is general,  to any dissipative system one can associate a differential form $\mathcal{G}$, defined in Eqn.(\ref{dynamicssystemdiffform}), from which a potential
function and  a non-potential function also called the residual part,  can be derived ($\omega_{exact}$ and $\omega_{antiexact}$). This generalizes the approach defined  by  Eqn.(\ref{Lyapunov2}).
It it happens that the associated form is closed, the system is a gradient or a gradient like  system. However, there are claims \cite{RalphChillEvaFasargova}, that 
every dissipative systems should be a gradient system. In the next section we duel on this claim. 
\
\section{A Generalized change of Variables}
As explained above, the  central objet of the discussions is the matrix of the partial derivatives $ \left [g_{j}(\vec{x}),\partial{x_{l}}\right]$.
If it is a symmetry matrix, the associated differential form to the dissipative systems  $\mathcal{G}$  is closed, and therefore there is a potential. 
It is not a symmetry matrix, there is a potential part and an additional form term  which is not exact. This state of affairs is invariant   within the change of variables  procedure , Eqn.(\ref{changeofvaribalestrans}). One has to resort to a non-linear kind of transformation as possible road map of remedy.  that is, 
    we want to create from one system of ODE, another one through this non-linear  transformation by maintaining the equivalence of the two. In this new ODE, the closed condition on an associated p=1 differential form is  imposed, and the solution 
derived from it, is used to find the potential.
We introduce a generalized change of variables $\phi^{\star\star}$(compare Eqn.(\ref{changeofvaribalestrans})): given a a function $\phi_{i}(y_{1},..,y_{N+1})$ $\in C^{n}(\vec{y})$, for our case of interest, $p=1$, and such that the inverse function  $\phi^{-1}(\vec{x})$ exist and ist differentiable:
\begin{widetext}
\begin{eqnarray}
\vec{x} \in U,  \omega \in \Omega^{n+1}_{p=1}(U,R), \hspace{0.3cm} \omega(x)=\sum_{j=1}^{N+1}  g_{j}(\vec{x})dx_{j}\in \mathcal{A}_{p=1}(R^{N+1},R),\hspace{1cm}\nonumber\\
\phi: U^{\prime} \rightarrow U,\hspace{0.2cm} \vec{y} \in U^{\prime},\hspace{0.4cm}x_{i}=\phi_{i}(y_{1},..,y_{N+1}));\hspace{4.5cm}\nonumber\\
\phi^{\ast}(\omega) \in \Omega^{n+1}_{p=1}(U^{\prime},R), \hspace{0.2cm} \phi^{\ast}(\omega)=\sum_{j=1}^{N+1}  g_{j}(\phi(\vec{y}))d\phi_{j}(y_{1},..,y_{N+1})=\sum_{j,i=1}^{N+1}  g_{j}(\phi(\vec{y}))\frac{d\phi_{j}(y_{1},..,y_{N+1})}{dy_{i}}{dy_{i}} \nonumber\\
\phi^{\star\star}(\omega)(\vec{y}):=\phi(\vec{g}))(\phi^{-1}(\vec{y}))=
\sum_{j}^{N+1}  \phi_{j}(g_{j})(\phi^{-1}(\vec{y})){dy_{j}} \hspace{2cm}\nonumber\\
\label{generalchangeofvaribalestrans}
\end{eqnarray}
\end{widetext}
i.e., $\phi^{\ast\ast}(\omega)$ is a $p=1$ form from $U^{\prime}\rightarrow \mathcal{A}_{p=1}(R^{N+1},R)$ in which the functions defining the canonical form are transformed by the function $\phi$ which also defines the coordinate transformation, in contrast to   $\phi^{\ast}(\omega)$, $\phi^{\ast\ast}(\omega)$ is nonlinear in $\Omega^{N+1}_{p=1}(R^{N+1},R)$.
With the previous definition we  transform Equ.(\ref{ODE})  to the new variables $\vec{y}$, without noise part \cite{Note6}:
\begin{widetext}
\begin{eqnarray}
\dot{x_{i}}=X_{i}(\vec{x})) \Rightarrow \dot{y_{i}}=Y_{i}(\vec{y})\hspace{3cm}\nonumber\\
\beta_{c}\ddot{y_{i}}+ \ddot{y_{i}}+f_{i}(\phi^{-1}(y_{1}),\ldots,\phi^{-1}(y_{N+1}))=0; \hspace{0.5cm} j=1,\ldots,N+1 \nonumber\\
f_{i}(\phi^{-1}(y_{1}),\ldots,\phi^{-1}(y_{N+1})):=  \phi_{i} ( g_{1}(\phi^{-1}\vec{y}), \ldots, g_{N+1}(\phi^{-1}\vec{y}))\nonumber\\
\mathcal{F}:=\sum_{i=1}^{N+1}f_{i}(\phi^{-1}(y_{1}),\ldots,\phi^{-1}(y_{N+1}))dy_{i}   \hspace{2.5cm}\nonumber\\
 \phi_{i}(\vec{y})=\sum_{j} \mathcal{D}_{i,j}(\vec{\mathbf{y}})y_{j}\hspace{4cm}
\label{ODEtransf}
\end{eqnarray}
\end{widetext}
$\mathcal{D}_{i,j}(\vec{\mathbf{y}})$ is a matrix function. Accordingly, it is taken to be invertible and such that the inverse of the function  $\mathcal{D}(\vec{\mathbf{y}})\vec{y}$ exist. 
When written as first order systems both Equ.(\ref{ODEtransf})($\dot{y_{i}}=Y_{i}(\vec{y})$)  and  Equ.(\ref{firstodersystem}) ($\dot{x_{i}}=X_{i}(\vec{x}))$
 should be equivalent under sufficient conditions for the function $\phi$, for example the Jacobian $\mathcal{J}=\det\mid\partial {y_{i}/\partial x_j}\mid$
 should be nonsingular. Then the following two differential  $p=r$ forms , $r=2(N+1)$,   $\Upsilon=(dx_{1}-X_{1}dt).....(x_{r}-X_{r}dt)$  and $\bar{\Upsilon}=(dy_{1}-Y_{1}dt).....(y_{r}-Y_{r}dt)$ are connected through the relation $\bar{\Upsilon}=\mathcal{J}\Upsilon$, and therefore
$\sum\partial{Y_{i}}/\nabla{y_{i}}=(1/\mathcal{J})(\sum\partial(\mathcal{J}X_{i}/\partial{x_i}))$, a relation that should be fulfilled 
for any given transformation\cite{HFlanders}. Therefore, with the p=1-form $\mathcal{F}$, defined in previous equation, we afford 
 the question  under what  general conditions one can find 
 a transformation $\phi$, such that  is closed form, i.e., $d\mathcal{F}=0$. This condition we
have seen already in Equ.(\ref{dynamicssystemdiffform}) and Equ.(\ref{dwZero}). For the $p=1$  form   $\mathcal{F}$, is given for the matrix of the partial derivatives.
From Eqns.(\ref{generalchangeofvaribalestrans},\ref{ODEtransf}), we have:
\begin{eqnarray}
\vec{x}=\mathcal{D}(\vec{y})\vec{y} \hspace{0.2cm} \Rightarrow \vec{y(\vec{x})}=\mathcal{D}^{-1}(\vec{y}(\vec{x})\vec{x} \hspace{0.1cm} \Rightarrow 
\frac{\partial g_{j}(\vec{y}_{j})}{\partial x_{k}}=\sum_{p}\frac{\partial g_{j}(\vec{y}_{j})}{\partial y_{p}}\frac{\partial (\vec{y}_{p})}{\partial x_{k}}=\nonumber\\
\sum_{p}\frac{\partial g_{j}(\vec{y}_{j})}{\partial y_{p}}
(\mathcal{D}^{-1}_{p,k}(\vec{{y}}(\vec{x}))+\sum_{l}\frac{\partial \mathcal{D}_{p,l}(\vec{{y}}(\vec{x}))}{\partial x_{p}}\cdot x_{l})\hspace{0.05cm} \Rightarrow\nonumber\\
\end{eqnarray}
\begin{eqnarray}
\left[\frac{\partial f_{i}(\vec{y})}{\partial x_{k}}\right]=
\left[\frac{\partial f_{i}(\vec{y})}{\partial x_{k}}\right]^{T}
=\left[\frac{\partial f_{k}(\vec{y})}{\partial x_{i}}\right]= \left(\sum_{j}\mathcal{D}_{i,j}(\vec{y}(\vec{x}))\hspace{0.05cm} 
\sum_{p}\frac{\partial g_{j}(\vec{y}(\vec{x}))}{\partial y_{p}}\hspace{0.05cm}\mathcal {D}^{-1}_{p,k}(\vec{y}(\vec{x}))\right)+\nonumber\\
\left(\sum_{j}\mathcal{D}_{i,j}(\vec{y}(\vec{x}))\sum_{p}\frac{\partial g_{j}(\vec{y}(\vec{x}))}{\partial y_{p}}\sum_{l} (\frac{\partial \mathcal{D}_{p,l}}{\partial x_{k}})
\cdot x_{l}\right)+
\left(\sum_{j}\frac{ \partial \mathcal{D}_{i,j}(\vec{y}(\vec{x}))}{\partial x_{k}} g_{j}(\vec{y}(\vec{x}))\right) \nonumber\\
\label{GeneralCondition}
\end{eqnarray}
Or equivalently with the definitions: $ g_{j}:= g_{j}(\vec{y}(\vec{x}))$,  $g_{j,\partial y_{k}}:=\frac{\partial g_{j}(\vec{y}(\vec{x}))}{\partial y_{k}}$, , and
  \hspace{0.1cm} $\mathcal{D}_{i,j,\partial x_{k}}:=\frac{ \partial \mathcal{D}_{i,j}(\vec{y}(\vec{x}))}{\partial x_{k}}$, and implicit sum over repeated indices :\nonumber\\
\begin{eqnarray}
\left(\left[\mathcal{D}_{i,j}g_{j,\partial y_{p}}\mathcal{D}^{-1}_{p,k}\right]-\left[\mathcal{D}_{i,j}g_{j,\partial y_{k}}\mathcal{D}^{-1}_{p,k}\right]^{T}\right)
+\hspace{4cm}\nonumber\\
\left(\left[\mathcal{D}_{i,j}g_{j,\partial y_{p}}\mathcal{D}_{p,l,\partial x_{K}}\cdot x_{l}\right]-\left[\mathcal{D}_{i,j}g_{j,\partial y_{p}}\mathcal{D}_{p,l,\partial x_{K}}\cdot x_{l}\right]^{T}\right)+\hspace{2cm}\nonumber\\
\left(\left[\mathcal{D}_{i,j,\partial x_{k}}\cdot g_{j}\right]-\left[\mathcal{D}_{i,j,\partial x_{k}}\cdot g_{j}\right]^{T}\right)=0\hspace{2cm} 
\nonumber\\
\label{MatrixEqtforMatrixD}
\end{eqnarray}
\nonumber\\
 The consistency condition is given by:
\begin{eqnarray}
k(\mathcal{F})(\vec{x})=\int\limits_{0}^{1}\mathcal{F}(\vec{tx};\vec{x})dt=\int\limits_{x_{0}=0}^{\vec{x}}\mathcal{F}(\vec{x})dx=
\sum^{N}_{i=1}\int\limits_{0}^{1}x_{i}f_{i}(t\vec{x})dt:
\hspace{0.2cm} \frac{\partial{k(\mathcal{F}))(\vec{x})}}{\partial{x_{i}}}=f_{i}(\vec{x})\nonumber\\
\label{DasdingansichforF}
\end{eqnarray}
 The nonlinear differential equation for $\mathcal{D}_{i,j}(\vec{y}(\vec{x}))$ subject to the consistency condition may be remarkably awkward for some systems as we have already checked
 , we treat therefore first the simplest cases where $\mathcal{D}_{i,j}(\vec{\mathbf{y}})$ is a constant matrix $\mathcal{D}_{i,j}$, i.e., applicable when the matrix of the partial derivatives $g_{j,\partial y_{k}}$ 
 is a constant matrix. In these cases only the first term in parenthesis  in Eqn.(\ref{MatrixEqtforMatrixD}) is non zero.
One obtains the following matrix equation:
\begin{widetext}
\begin{eqnarray}
x_{i}:=\sum_{j=1}^{N+1} \mathcal{D}_{i,j}y_{j}, \hspace{0.2cm} f_{i}(\phi^{-1}(\vec{x})):=\sum_{j=1}^{N+1} \mathcal{D}_{i,j}g_{j}(\phi^{-1}(\vec{x})))\nonumber\\
\nonumber\\
Equ.(\ref{dynamicssystemdiffform}), Equ.(\ref{MatrixEqtforMatrixD}) \Rightarrow \mathcal{D}^{T}\mathcal{D}[g,\partial{y}]=[g,\partial{y}]^{T}\mathcal{D}^{T}\mathcal{D}
\label{Matrixequation}
\end{eqnarray}
\end{widetext}
If the form $\mathcal{G}$ where closed, then the solution is the identity matrix $\mathcal{D}=I$.
The last equation constitutes  a necessary condition to be fulfilled by the solution  of $[g,\partial{y}]$ as a function  of $\mathcal{D}$, which is obtained  from the Eqn.(\ref{Dasdingansich}). Therefore, with the consistency condition  Eqn.(\ref{DasdingansichforF}) one obtains: 
\nonumber\\
\begin{equation}
k(\omega)(\vec{x})=\int\limits_{0}^{1}\omega(\vec{tx};\vec{x})dt=\int\limits_{x_{0}=0}^{\vec{x(y)}}\omega(\vec{x})dx \Rightarrow 
 \mathcal{D}(\mathcal{D}^{T})^{-1}=[g,\partial{y}]^{T}
\label{MatrixequationSimple}
\end{equation}
\nonumber\\
\nonumber\\
The solution fulfills  the necessary condition Eqn.(\ref{Matrixequation}). As the matrix $[g,\partial{y}]$ is constant,  a  straightforward way to get the  last relation is to use the ansatz for the potential $U=\sum_{i}\frac{1}{2}\phi_{i}(\vec{x})^{2}$ with $\phi_{i}(\vec{x})=f_{i}(\vec{x})$ and to apply on $U$ the d-operator  and then dd.
\\
\section{Application of the generalized change of variables concept.}
\subsection{Two-dimensional Josephson Junctions Arrays:}
  The resistive behavior of a two dimensional Josephson array with a given  frustration parameter (the ratio of the perpendicular magnetic field to the flux quantum per plaquette ($f=\frac{M}{N}[\Phi_{0}]$, where $M,N$ are natural numbers , with $\Phi_{0}=hc/2e$ ), have been a matter of intense research \cite{RafaelRangelI}\cite{PorterStroud}\cite{MartinoliLeemann}, see page 21 in \cite{FaziovanderZant}.
In this case, primary there are phase variables  for which there are  functions $\Phi_{k}(y)$ which  express their dependence  on the variables $y$, and an auxiliary matrix   $\omega$. Then  we have for any $(M,N)$ for the functions appearing in  Eqn.(\ref{ODE}) are  given  by  Eq.(3) in  \cite{RafaelRangelI}, 
where $\forall j, \beta_{cj}=\beta_{c}$. Fist, accordingly,  one associates a differential form to the dissipative system and   checks the cross derivatives to  find  that $\frac{\partial g_{j}(\vec{y}(\vec{x}))}{\partial y_{k}}\ne \frac{\partial g_{k}(\vec{y}(\vec{x}))}{\partial y_{j}}$, i.e., the associated diff-form is not closed.
Then the second step is carried  out and  we obtain with $\vec{x}=\mathcal{D}\vec{y}$, and $a_{ik}:=\frac{1}{2}\sum_{j=1}^{N+1}D_{ij}\omega_{jk}$, the expression for the $f_{i}(\vec{x})$(Eqt.(5) in \cite{RafaelRangelI}),
 and so finally the specific equation for the matrix $\mathcal{D}$ is derived from Equ.(\ref{Matrixequation}):
\begin{widetext}
\begin{eqnarray}
\mathcal{D}^{T}\mathcal{D}[\omega\Phi_{k,\partial{y}}]=[\omega\Phi_{k,\partial{y}}]^{T}\mathcal{D}^{T}\mathcal{D}\hspace{0.2cm} \Rightarrow 
\frac{1}{2}\mathcal{D}^{T}\mathcal{D}=\Phi_{\partial{y},k} \omega.
\label{MatrixequationTDJJA}
\end{eqnarray}
\end{widetext}
With the help of Equ.(\ref{MatrixequationTDJJA}), the potential can be calculated in principle for any $M,N$ (see Eqn.(6) and Eqn.(7) in \cite{RafaelRangelI}),
 and in particular,  we have calculated the potential for two important  cases, namely, $f=\frac{1}{2},\frac{1}{3}$ \cite{RafaelRangelI},\cite{RafaelRangelNr2}.
The authors in  \cite{YongChoiNr1}, \cite{YongChoiNr2},\cite{YongChoiNr3}  among many others , hail the importance of these two cases. There are noise terms associated with these systems, and is relevant  to say that theses terms transforms differently in the new systems of coordinates defined by the matrix $\mathcal{D}$, such that, the intensity of the noise is bigger for  some variables (for details for $f=\frac{1}{2}$ see \cite{RafaelRangelI}). This a matter with physical consequences when one consider the whole stochastic behavior of the system (this issue is under study).
 Furthermore, they find the case $f=\frac{2}{5}$  is for some unknowns reasons a case that emulate in a universal form cases near incommensurability where $f=(3-\sqrt{5})/2$. The matrix $\mathcal{D}$  for $f=\frac{2}{5}$   is a complicated object versions of which we have already derived. However, a static disorder due to a non-vanishing variance in the Josephson parameters makes the SDE also a quenched disordered system\cite{FilatrellaPedersenWiesenfeld}. Summarizing, the dynamic equations for two-dimensional Josephson  arrays under a perpendicular magnetic field are gradient like systems with the particularity that  the potential   are not bounded from below being tilted washboard functions. 

\subsection{Examples with Josephson Junctions Circuits.}
There are many models  of Josephson circuits \cite{MichaelTinkham}.  We use Equ.(\ref{Matrixequation}) to obtain the potential  for the Josephson junction  circuit defined in \cite{ChateManneville}. We write their equations $(2.4-2.7)$ in the form of Equ.(\ref{ODE}) where thermal noise is neglected for the variables $\vec{y}=(y_{1}=y,y_{2}=\delta,y_{3}=\zeta)$, 
and the three functions $\vec{g}=g_{1}(\vec{y}),g_{2}(\vec{y}),g_{3}(\vec{y})$. \\
\begin{eqnarray}
\beta_{c}\ddot {\delta}=-r\dot{\delta} +(\imath-\sin{\delta} -\zeta)\nonumber\\
\frac{1}{\beta_{L}}\dot{\zeta}=-\zeta +y\hspace{2.450cm}\nonumber\\
\label{JosephsonJunctioncircuit}
\end{eqnarray}
The last equations give rise to a dissipative system in the form of Equ.(\ref{firstodersystem}):\\
\\
\\
\begin{eqnarray}
\hspace{3cm}\dot{\delta}=y=:g_{1}(y)\hspace{5.3cm}\nonumber\\
\dot{y}=-\frac{r}{\beta_{c}}y+  \frac{1}{\beta_{c}} (\imath-\sin{\delta} -\zeta)=:g_{2}(y,\delta,\zeta)\hspace{1.1cm}\nonumber\\
\dot{\zeta}=-\frac{1}{\beta_{L}}\zeta +\frac{1}{\beta_{L}}y=:g_{3}(y,\zeta)\hspace{3.25cm}\nonumber\\
\label{JosephsonJunctioncircuitNr2}
\end{eqnarray}
If we can find a potential  it constitutes  a gradient  system. One first verifies that $\partial g_{i}(\vec{\mathbf{y}})/\partial \mathbf{y}_{j} \neq \partial g_{j}(\vec{\mathbf{y}})/\partial \mathbf{y}_{i}$, $1\leq i,j \le  3$, and therefore,  for  the associated $p=1$ differential form  $\mathcal{G}(\vec{\mathbf{\mathbf{y}}})$(Equ.(\ref{dynamicssystemdiffform}))
 $d\mathcal{G}(\vec{\mathbf{\mathbf{y}}})\not=0$, and therefore, there  is not a potential  in the sense of Equ.(\ref{Lyapunov1},\ref{Dasdingansich}).
  In this form, the Josephson junction circuit was numerically simulated for particular values of the parameters $(\imath,r,\beta_{c},\beta_{L})$\cite{ChateManneville}. 
We find the potential in two steps:
First we study the linear case for small value of $\delta << 1$, and obtain accordingly previous explanations the following  matrix equation:
\begin{equation}
 \mathcal{D}(\mathcal{D}^{-1})^{T} =[g_{i}(\vec{y}),\partial{y_{j}}]^{T}
\label{MatrixconditionforJosephsonCircuits}
\end{equation}
We then obtain the following matrix equation for the case under study, i.e., for the $ 3X3$ matrix $ \mathcal{D}$: \nonumber\\
\\

$[g_{i}(\vec{y}),\partial{y_{j}}]=
\left( \begin{array} {llcl} 
1 & 0 & 0\\
\frac{-r}{\beta_{c}}& \frac{-1}{\beta_{c}} & \frac{-1}{\beta{c}}\\
\frac{1}{\beta{L}} & 0  & \frac{-1}{\beta_{L}}\\
\end{array} \right)=
 \mathcal{D}^{T}(\mathcal{D}^{-1}) $\\
\\
\\
 This matrix equations defines $9$ polynomial relations  in each case for the $9$ constituents components $d_{ij}$ of $\mathcal{D}$, $ 1 \le i \le 3, 1\le j \le 1$. No analytical solution seems achievable, and only a systematic numerical study  is possible.
\nonumber\\
From Eqn.(\ref{dynamicssystemdiffform}) the following relations holds for  $(\phi(\vec{y})=\mathcal{D})$: \nonumber\\
\begin{equation}
\forall i \ne j, \hspace{0.2cm} \frac{\partial f_{i}(\mathcal{D}^{-1}(\vec{x}))}{\partial x_{j}}=\frac{\partial f_{j}(\mathcal{D}^{-1}(\vec{x}))}{\partial x_{i}}
\label{equalityCroosDerivaties}
\end{equation}
and therefore, for  p=1-diff-form $\mathcal{F}:=\sum_{i=1}^{N+1}f_{i}(\mathcal{D}^{-1}(x_{1}),\ldots,\mathcal{D}^{-1}(x_{N+1}))dx_{i}$  is closed and exact.
The potential is $k(\mathcal{F}(\vec{x}))=\int\limits_{x_{0}=0}^{\vec{x}}\mathcal{F}(\vec{x})dx$. After that, we put a $-cos(\delta)$ in place of the variable $\delta$ alone.
Notwithstanding, the road map  is to apply the procedure given   in Eqn.(\ref{MatrixEqtforMatrixD}) and 
 Eqn.(\ref{DasdingansichforF}).
In a similar way, the model of the circuit \cite{FangfangZhang} can be analyzed, and is actually  a matter under study.
\nonumber\\
\section{Summary and Perspectives.}
In this work, two relevant results are obtained. On the one hand,  from the theory of differential forms, we show that a   p=1 differential form can be associated with  a dissipative systems. This diff-form may be closed or not. If it is a closed form then we can associate to the dissipative system a potential. it constitutes a gradient system or a gradient like system.  If it  is no close, (Eqn.(\ref{dynamicssystemdiffform}),  there exist  no  coordinate transformation that could change this property. The  property  of been non-closed is invariant under coordinate transformation. Through the introduction of the Homotopic  or k-operator, in combination with the d-operator,
 in either   case, one can decompose the diff-form in an exact pat and an antiexact part. This decomposition is universal valid, and express that to any dissipative systems, its dynamics is given by a potential like part and a non-potential term.
On the other hand,  
 If it is not a closed form, we proposed in this work ,  a nonlinear matching procedure by  "rotating" both, the vector  function $\mathcal{G}(\vec{y})$ and the coordinate system $\vec{y}$, simultaneously under the transformation $\phi(\vec{y})$ and imposing for the new p=1 diff-form the necessary condition for being closed. 
 This is the generalized change of variables (Eqn.(\ref{generalchangeofvaribalestrans}) which we implement through the introduction of a matrix function (Eqn.(\ref{ODEtransf}). For the p=1 diff-form constructed (Eqn.(\ref{ODEtransf}), we derive an equation (Eqn.(\ref{MatrixEqtforMatrixD})) whose results give
the equation defining this transformation, together with the consistency condition (Eqn.(\ref{DasdingansichforF}).  It constitutes our main result.  We claim that the same conditions that guaranties the existence  of a solution of the dynamical systems (Eqn.(\ref{firstodersystem})), also guaranties  the solution a  Eqn.(\ref{MatrixEqtforMatrixD}).
For some cases, the transformation  can be defined through a constant matrix, and an analytical solution can be found, i.e., the case of two dimensional Josephson arrays under a magnetic field \cite{RafaelRangelI}, \cite{RafaelRangelNr2}. 
For the cases in \cite{FangfangZhang}, and \cite{ChateManneville}, it appears to us that only a numerical study is feasible.
These results bring us to the main purpose  of this work, namely, to answer the question, is a dissipative systems always a gradient system ?.
The examples discussed above, show certainly, that whereas the differential form formalism say there is not a potential, our approach finds that indeed it  exist,
in the context of the generalized change of variable concept.  For the time being, we can only  say for sure, that for the kind of systems where 
Eqt.(\ref{MatrixequationTDJJA}) apply, one can find a potential, i.e., the two-dimensional Josephson  arrays.  Also, for systems where Eqt.(\ref{MatrixconditionforJosephsonCircuits}) applies, i.e., Josephson junction circuits,  although the resulting algebraic equations appears to be accessible only through   pure numerical simulations.
 And in general, depending on each particular system, one has to figure out a strategy for find the generalized transformation defined in 
 Then we have the natural question: given a particular dissipative systems  there exist such a transformation $\phi (\vec{y})$, as defined in Eqn.(\ref{ODEtransf}),
  , and such that it satisfies  Eqn.(\ref{GeneralCondition}),  Eqn.(\ref{MatrixEqtforMatrixD}), and  Eqn.(\ref{DasdingansichforF}). This procedure may be a difficult one,  but even in principle, all dissipative systems are gradient systems , or gradient like systems.
Our work naturally conduct us  to consider dissipative systems defined in infinite  dimensional Bach spaces, for example,  Navier-Stokes equations \cite{WolfvonWahl}, or the Generalized Time-dependent 
Ginzburg-Landau equations (GTDGLE)\cite{KramerWattstobinKrahenbul},\cite{ThesisRafaelRangel},\cite{ArmenGulian}.
This is an extended systems represented by an evolution equation given by an abstract  parabolic partial differential equations.  The corresponding Banach space is of  infinite  dimension,  and the associated operators are  treated in a Hilbert space with a combination of the concepts of differential forms and functional analysis. In analogy with finite dimensional dissipative systems\cite{GrahamTelNr2}, for example the GTDGLE posses attractors with fractal dimension in Banach Spaces\cite{PRA-RafaelRangel}.
This matter is under investigation.
Summarizing, our approach to the problem to the existing of a potential for a dissipative dynamical systems 
provides a clear mathematical set up for finding the potential, and confirms the equality of the class of dissipative systems and the class of gradient systems.
\section{Apendix}
\subsection{Discussion of  Graham Theory in the light of  Differential forms.}
In Graham scheme the dissipative model is given from the upset as in Eqn.(\ref{OrthogonalityCondition}). Their  claim  is referred to the existence of a potential with respect to a 
a certain given positive semi-definite matrix $\mathcal{S}$. In the context of this   particular class of  continuous dissipative systems , they showed that the existence of the potential is not a generic property.
 It consist in a decomposition (assuming possible) of a given function that govern the
dynamics, i.e., $\dot{\vec{\tilde{x}}}=f(\vec{x})$,   in the form:  $ f(\vec{x})=\mathbf{D}\nabla \mathbf{V}=(\mathbf{D}^{symm} + \mathbf{D}^{skew})\nabla \mathbf{V}$ in a part proportional to a positive definite  symmetry matrix and another proportional to an skew symmetry one. In this way,  the residual part equals a rotation of a gradient of a function, i.e.,  $\nabla \mathbf{V}$,  which is given by 
$\vec{v}=\kappa \Omega \nabla \mathbf{V}$, $\Omega$  is the unit skew-symmetry, $\kappa$ a set of parameters, which guaranties  the orthogonality condition is fulfilled.
Accordingly, Graham obtains\cite{WioHSDezaRR} (see Equ.(\ref{OrthogonalityCondition})): \\
 $\vec{\tilde {f}}:= \nabla \mathbf{V}=-(S-k\Omega)^{-1} \vec{f}$, which represents a rotation of the originally function in Equ.(\ref{OrthogonalityCondition}) by the matrix $\mathbf{D}^{-1}=-(S-k\Omega)^{-1}$. Then, in the new coordinates $\vec{\tilde{x}}$,  the equation:   $\dot{\vec{\tilde{x}}}=-(S-k\Omega)\vec{{f}}(\vec{\tilde{x}})$, with $\vec{\tilde{x}}=-(S-k\Omega)^{-1} \vec{x}$.
   Therefore, the existence of a potential  requires the equality of the cross derivatives for $\vec{\tilde {f}}$,  and this precisely the condition they call the integrability condition for $\mathbf{V}$. 
   This is the Graham scheme  remedy for the lack of the equality of the cross derivative for function $f(\vec{x})$. 
 This requires the existence of an appropriate set of parameters such that the integrability condition can be fulfilled\cite{WioHSDezaRR}.  But how to know if these parameters exist?.
The answer within the differential form math formalism is to check if  an integrating factor exist  for the  dynamics system defined just by
the first line of Equ.(\ref{Lyapunov2}), i.e.,   for the $p=1$ form  $\omega:=\vec{f}d\vec{x}$, is  there a function $h$, such that  $\left(d(h\omega)\right)=\left(d(h)\wedge w + h d\omega\right)=0$?. Therefore,  the matrix  $-(S-k\Omega)^{-1}$ represents the integrating factor for the form $\omega$. 
The condition $\omega \wedge d(h\omega)=\omega\wedge ( d(h)\wedge w + \omega \wedge h d\omega)=0$, $\Rightarrow$  $\left [\omega \wedge d\omega=0\right]$ , which should then be fulfilled  as a necessary and sufficient   condition  for existence of the integrating factor function h(x), no necessary a constant matrix function \cite{HCartanII}.
\begin{eqnarray}
\omega \wedge d\omega=(\sum^{N}_{l=1}f_{l}dx_{l}) \wedge (\sum^{N}_{k,i=1}\frac{\partial f_{i}}{\partial x_{k}}dx_{k}dx_{i})=\sum^{N}_{l,k,i=1}f_{l}\frac{\partial f_{i}}{\partial x_{k}}dx_{l}dx_{k}dx_{i})=0,
\label{factorexistencecondition}
\end{eqnarray}
which means that the vector function $\vec{f}$ is orthogonal to its $"rotor"$. In fact, previous  relation should be checked before getting involved in the Graham  scheme. In all the examples where the scheme was successfully applied \cite{DezaRRWioHS}, previous condition must have been  satisfied. In other words, the Graham scheme applies for the special class of functions   where  the p=1 diff-form  defined in Eqn.(\ref{dynamicssystemdiffform}) satisfies Eqn.(\ref{factorexistencecondition}).
\section{ Acknowledgments.}
We  thank Prof. Dr. Thomas Klinker, Universitaet Hamburg, Germany,  for donating  the important book for this work on Differential Forms by Henry Cartan and for
 financial support for this project. I thank Dr. Hector Giusti , DIC.USB, for donating a pc, which was used for writing this work.



\begin{thebibliography}{XXX}
\bibitem{MaxwelleqtsExample} The Maxwell equations and the differential forms formalism. The equations  can be written in $\mathcal{R}^{4}$, where an $\vec{\tilde{x}}$ $\in \mathcal{R}^{4}$ has coordinates $(x^{0}=t, (x^{1},x^{2},x^{3}=\vec{x}))$,  through  the introduction of 
a p=2 diffrential form: $F=\frac{1}{2}=F_{\alpha,\beta}dx^{\alpha}dx^{\beta}$, where $F_{\alpha,\beta}$ is the field-strength tensor. The homogeneous  Maxwell equations are the coordinate free stament that the form $F$ is closed, i.e., 
$dF=0$. It is also exact and therfore ther exist  a p=1 diffential form, the potential, $A=A^{\alpha}dx^{\alpha}$, , such that 
$dA=\partial_{\beta} A^{\alpha}-\partial_{\alpha} A^{\beta}=F$, threfore from Eqt.(\ref{Dasdingansich}), \hspace{0.2cm} $A=k(F)(\vec{\tilde{x}})=\left(\int\limits_{0}^{1}F(t\vec{\tilde{x}})dt\right)\vec{\tilde{x}}$. The inhomogenous Maxell equation can be writen as a  combination of the the d-operator and the *-operator as $d*F=-*J$,  where the p=1 diff-form  $J=-\rho dt + \vec{J}.dx$ is written in terms of the charge density and the current density respectively. The consistency equation $dd*JF=0$ implies $d*J=0$, which express concservation of charge. This work  does not make use of the *-operator. See references \cite{HFlanders,WThirringI} and chapter 1.3 in \cite{WThirringII}.
\end{thebibliography}
\end{document}